\documentclass[aps,prd,article,showpacs]{revtex4}%
\usepackage{amsmath}
\usepackage{graphicx}
\usepackage{epsfig}
\usepackage{amsfonts}
\usepackage{amssymb}%
\providecommand{\U}[1]{\protect\rule{.1in}{.1in}}
\providecommand{\U}[1]{\protect\rule{.1in}{.1in}}
\begin{document}
\title{\ How the active and diffusional nature of brain tissues \\can generate monopole signals at micrometer sized measures}
\author{Alejandro Cabo$^{*}$ and Jorge J. Riera$^{**}$ \bigskip}

\address{$^*$ Department of  Theoretical Physics, Instituto de Cibern\'etica, Matem\'atica y F\'isica ICIMAF, La Habana, Cuba. \bigskip\\
$^{**}$ Department of Biomedical Engineering, Florida International University, Miami, Florida.\bigskip }
\begin{abstract}
\noindent We investigate mechanisms which could generate transient monopole
signals in measuring current source density (CSD), as it had been indicated to
occur in recent small volume experiments. A simple model is defined for this
purpose. It is emphasized that the active nature of the neural biological
activity, with its ability to generate ionic density imbalances, might be able
to induce appreciable monopole signals in CSD detectors at micrometer scales.
Thus, it follows that when both diffusive and ohmic transport are considered
to be present in neural tissues, potential measures in micrometer regions can
include appreciable electric monopole signals, for sufficiently small values
of the ratio $\frac{\sigma a^{2}}{\epsilon D}$, where, $\sigma$ is the
conductivity,  $\epsilon$ is the dielectric constant, $D$ is the diffusion constant and $a$ is the linear dimension of
the ionic charge densities generated by the neural processes. Ranges of
possible magnitudes for these parameters in the considered experimental
studies are estimated. The analysis indicates values for the ratio between the
dipolar and monopole signals which are close to the ones measured in Pyramidal
cells in recent experiments. The measured results for Spiny Stellate cells are
also qualitatively described by the model by predicting a finite monopole
signal in combination with vanishing dipolar and quadrupole ones.

\end{abstract}

\maketitle

\section{\bigskip Introduction}

The finding by Riera et al. in references \cite{jriera,jriera2} about the
possible existence of temporal unbalance in the extracellular ionic charge
have generated a very intense theoretical debate with position divided in
those that support it (See \cite{Bedard&Destexhe,jriera2}) and those that are
opposed to it(See \cite{Gratiy}). If monopole current sources can coexist
transiently with dipolar ones at small physical scales, this coexistence may
have a profound implication in the way we interpret current small-scale EEG
and MEG data as well as in the models we employ today to simulate this type of
brain data recordings.

In the present work we intend to argue that the inclusion of diffusion
in the transport properties of the brain tissues, in addition to
the consideration of the active properties of the membrane processes, could
describe the above mentioned recent and surprising findings: namely, the
evidences of non vanishing monopole components in the evoked potentials
measured in mesoscopic portions of the brain tissues \cite{jriera,jriera2}. It
should be mentioned that the relevance of diffusion and other processes for
the correct explanation of modern neurophysiology measurements had been
discussed in detail in reference \cite{Bedard&Destexhe}.

We start the presentation by reviewing the standard quasistatic approximation
for the description of brain tissues, in which they are considered as perfect
conductors \cite{nunez}. Afterwards, the simple model in which the work is
based will be defined. The main addition to the Ohmic conduction property of
the brain tissues is the inclusion of the possibility of diffusion of the
charges. The model assumes a single ionic component of the diffusion in order
to simplify this starting analysis. The general Poisson equation describing
the electric potential at the interior of the tissue medium is written. More
complex situations can be treated with the help of the general discussions of
diffusion effects in the literature (See \cite{diffusion}).

Next, the equations are considered for simple one dimensional (1D) models. A
1D model is constructed in which the cell interior is assumed to be the
negative $x$ axis, and the Ohmic and diffusional tissue fills the $x>0$
region. A cell membrane was assumed to lay at the $x=0$ point of the one
dimensional model, which during some time interval was allowing the flow of
positive ions to the $x<0$ cell's interior region. After that, the membrane
was assumed to be blocked. Then, the equations for this 1D model were solved
for the potential and densities as functions of the time and spacial variable
$x$. The solutions were found for two values for the single relevant parameter
defining them $\sigma^{\ast}=\frac{\sigma a^{2}}{\epsilon D}$, where $\sigma$
is the Ohmic conductivity of the tissue, $a$ is the above defined linear
length of the region being the support of the negative charges compensating
the net ionic charges trapped within the cell interior. Further, $\varepsilon$
is the dielectric constant of the tissue and $D$ is the diffusion constant of
the medium.

The 1D equations are exactly solved as Fourier integrals determined by the
initial conditions. The solutions are firstly discussed for the zero
conductivity limit value, in order to examine the case in which only diffusive
transport is allowed. The results clearly illustrate how the diffusion is
able to dissipate the initial cloud of negative charges located around the
cells, by translating those charges to infinity, and then leaving the
electric field generated by the charges inside the cell unscreened. Then, it
follows that in this extreme unphysical limit, monopole charges can be
"measured" at arbitrary faraway regions from the considered cell model. Next,
a little more realistic family of solutions is discussed for a finite value of the
unique parameter $\sigma^{\ast}$. The results show that in this case, the
initial cloud of charges also starts to be dissipated by the combined
conduction and diffusion currents, but now leading the negative charges to
conform a distribution which stabilizes a 1D static Yukawa solution. At this
configuration, the system stops to dissipate energy, because the total current
vanishes at all the spatial points. \ It follows that the potential in the
faraway regions always grows with the time, up to the value corresponding to
the static Yukawa solution. However, this static configuration  is clear
that can not remain established for a long time, due to an additional physical
process, not being taken into account in the present model. That is the slower,
diffusion processes across the membrane, which tends to return the positive
charges trapped within the cell to the tissue medium surrounding it.
Therefore, the value of the Yukawa potential at some measuring electrode point
can be considered as an estimate of the maximal signal arriving to this
electrode from a firing cell. Then, we will adopt this potential values as
giving a measure of these firing processes at the instant of their maximum.
It should be noted, that this picture furnishes a qualitative explanation for
the relevant times variations in the order of milliseconds shown by the CSD process
in neural systems.  This is the time scale of the membrane firings in the neurons
and  the consecutive ions return processes to the extra cellular medium.

Next, it is argued that all the determined solutions of the 1D problem lead to
corresponding solutions of the spherically symmetric 3D problem in which the
radial distance $r$ coincides with the $x$ coordinate of the 1D problem.
Further, the 3D potential and charge densities are equal to the 1D solutions
after divided by $r$. In this mapping process the 1D static Yukawa solutions
lead to corresponding 3D Yukawa field distributions. In these solutions
 the central point charge,  models the positive ions being
collected inside the cell after the diffusion through the membrane. The
screening length associated with the Yukawa solution is defined by the square root
 of the above defined constant $\sigma^{\ast}.$

The work continues by considering two models. The first  one is for the firing of a
spherically symmetric cells. It is basically described by the Yukawa
solution at an instant in which the signal at the electrodes approaches its
maximal value. The other one, is defined by a superposition of two spherically
symmetric solutions, each of which is also described by Yukawa fields but
corresponding to opposite charges which are  displaced in a vector
$\overrightarrow{l}$,  corresponding to a similar length as the constant $a$. This is considered
in the work as a simplified model for typical non-spherical CSD
configurations. The Pyramidal cell action potential, and its physical
description is reported in reference \cite{jriera} and shows a more complex
structure in which quadrupole components are also measured. In those
experiments they also exhibit monopole and dipole components. The measured
ratio between the dipolar and the monopolar signals will be taken here as a
reference for a qualitative comparison between the predictions of the proposed
model and the experimental results.

The field configurations are employed for estimating the ratio between the
potential at the electrodes created by the dipolar, $\phi_{d}(r)$, and
monopolar, $\phi_{m}(r)$, components of the Pyramidal cell action potentials.
The Spiny Stellate cells, in which CSD  also demonstrate weak but non vanishing
monopolar components as reported in \cite{jriera}. The fact that the dipolar
and quadrupolar components both vanish, can be estimated as  indicating that
the model reasonably describes the measurements associated to the Spiny
Stellate neurons in that work. The range of values of the relevant parameter
in cortical systems $\sigma^{\ast}=\frac{\sigma a^{2}}{\epsilon D}$ were estimated after
considering the reported values for the brain tissue conductivities, the
spatial dimensions $a$ and their typical separations $l$ of the charge
densities in the action potentials of Pyramidal cells measured in
\cite{jriera}.  The reported values of the diffusion constants in extra cellular media
 for the $N_a$  and $C_a$ ions, and the permeability values for the cortical
 tissues  were  employed.

The results allow for the extraction of a main conclusion of this work: for the
experimental conditions reported in \cite{jriera,jriera2}, the presence of
the surprising monopolar components in the estimated evoked potentials in
mesoscopic regions of the brain tissues, can be justified after considering
the active nature of the cell membrane processes in combination with the
presence of diffusion in the intercellular brain medium.

It should mentioned that upon writing this article, we had been informed about
the appearance of a work which independently underlines the relevance of the
active and diffusional characters of the neural processes in the brain
\cite{neuron}.

The exposition proceeds as follows: the basic ideas of the quasistatic model
for the neural tissues are shortly reviewed in Section 2. In Section 3 the
model for the transport equations including diffusion is introduced.  Section 4
considers the solution of the equations for the density and the potentials,
in the case of the simpler one dimensional systems. Section 5 is devoted to
construct 3D solutions from the 1D solutions. The simple models for cell action potentials
showing a spherically symmetric monopole and a cylindrical symmetric dipolar
CSD are presented in Section 6.  The estimates of the relative strengths
of the two configurations, and the \ range for the experimental values for the
parameter $\sigma^{\ast}$ are also described there.

\section{The \ quasistatic model for the brain tissue}

In this initial section, we will
review the elements of the volume conductor model for brain tissues. Let us
consider a conductive tissue medium that can efficiently screen
electric charge densities generated within it by the neurological system.
The current and charge distributions associated with the conductive tissue
will be, on one side, the Ohm conductivity current component
$\overrightarrow{J}(\overrightarrow{x},t)=\sigma(\overrightarrow{x}%
)\overrightarrow{E}(\overrightarrow{x},t)$, and its associated charge density
function $\varrho(\overrightarrow{x},t)$. In addition, the system will exhibit
the so called impressed current and charge densities, generated by complex
biological processes $\ \overrightarrow{J}_{imp}(\overrightarrow{x}.t)$ and
$\varrho_{imp}(\overrightarrow{x},t).$ \ The total current and charge
densities in the conductor tissue will be defined as
\begin{align}
\varrho_{T}(\overrightarrow{x},t) &  =\varrho(\overrightarrow{x}%
,t)+\varrho_{imp}(\overrightarrow{x},t),\\
\overrightarrow{J}_{T}(\overrightarrow{x}.t) &  =\overrightarrow{J}%
(\overrightarrow{x},t)+\overrightarrow{J}_{imp}(\overrightarrow{x}%
.t).\nonumber
\end{align}
\ The displacement field is given in the form%
\[
\overrightarrow{D}(\overrightarrow{x},t)=\epsilon\overrightarrow{E}%
(\overrightarrow{x},t),
\]
in which the dielectric properties are assumed to be space independent.
\ Thus, the Maxwell equations become
\begin{align}
\overrightarrow{\nabla}\times\overrightarrow{B}(\overrightarrow{x},t) &
=\overrightarrow{J}_{T}(\overrightarrow{x}.t)+\frac{\partial}{\partial
t}\overrightarrow{D}(\overrightarrow{x},t),\text{ \ }\overrightarrow{\nabla
}\times\overrightarrow{E}(\overrightarrow{x},t)=-\frac{\partial}{\partial
t}\overrightarrow{B}(\overrightarrow{x},t),\\
\overrightarrow{\nabla}\cdot\overrightarrow{D}(\overrightarrow{x},t) &
=\varrho_{T}(\overrightarrow{x},t),\\
0 &  =\overrightarrow{\nabla}\cdot\overrightarrow{J}_{T}(\overrightarrow{x}%
.t)+\frac{\partial}{\partial t}\overrightarrow{\nabla}\cdot\overrightarrow{D}%
(\overrightarrow{x},t)=\overrightarrow{\nabla}\cdot\overrightarrow{J}%
_{T}(\overrightarrow{x}.t)+\frac{\partial}{\partial t}\varrho_{T}%
(\overrightarrow{x},t)\nonumber\\
&  =\overrightarrow{\nabla}\cdot\overrightarrow{J}(\overrightarrow{x}%
.t)+\frac{\partial}{\partial t}\varrho(\overrightarrow{x},t)+\nonumber\\
&  \text{ \ \ \ \ }\overrightarrow{\nabla}\cdot\overrightarrow{J}%
_{imp}(\overrightarrow{x}.t)+\frac{\partial}{\partial t}\varrho_{imp}%
(\overrightarrow{x},t).
\end{align}

We will assume that both: the conductor and the current densities
independently satisfy the local charge conservation condition. That is%
\begin{align}
\overrightarrow{\nabla}\cdot\overrightarrow{J}(\overrightarrow{x}%
.t)+\frac{\partial}{\partial t}\varrho(\overrightarrow{x},t)=0,  &  \text{ }\\
\overrightarrow{\nabla}\cdot\overrightarrow{J}_{imp}(\overrightarrow{x}%
.t)+\frac{\partial}{\partial t}\varrho_{imp}(\overrightarrow{x},t)  &  =0.
\end{align}

Let us consider now a small region outside the neural system, that is, being
inside the conductor tissue and assume that the conductivity, is a constant
parameter. If it is also assumed that the impressed currents and charges
vanish in this  region, the charge conservation condition can be written as
\begin{align}
0 &  =\overrightarrow{\nabla}\cdot\overrightarrow{J}(\overrightarrow{x}%
.t)+\frac{\partial}{\partial t}\varrho(\overrightarrow{x},t)\nonumber\\
&  =\frac{\sigma}{\epsilon}\overrightarrow{\nabla}\cdot\overrightarrow{D}%
(\overrightarrow{x}.t)+\frac{\partial}{\partial t}\varrho(\overrightarrow{x}%
,t)\nonumber\\
&  =\frac{\sigma}{\epsilon}\varrho(\overrightarrow{x},t)+\frac{\partial
}{\partial t}\varrho(\overrightarrow{x},t)\nonumber\\
&  =(\frac{\sigma}{\epsilon}+\frac{\partial}{\partial t})\varrho
(\overrightarrow{x},t).
\end{align}

Thus, the tissue conductor charge density (working within the assumptions of
this work) satisfies the simple equation
\begin{equation}
(\frac{\sigma}{\epsilon}+\frac{\partial}{\partial t})\varrho
(\overrightarrow{x},t)=0,
\end{equation}
with the following solution:
\begin{equation}
\varrho(\overrightarrow{x},t)=c\exp(-\frac{\sigma}{\epsilon}t).
\end{equation}

In other words, if at a given time, the charge density in the tissue was
forced to be non zero, it would decay with the time constant%
\begin{equation}
\tau=\frac{\epsilon}{\sigma}.
\end{equation}

For the cortex conductor tissue, the decay time is on the order of
$\ 1.66\times10^{-11}$ $\frac{\epsilon(f)}{\epsilon_{0}}$seconds (See
reference \cite{nunez}), \ where $\frac{\epsilon(f)}{\epsilon_{0}}$ \ is the
frequency $f$\ dependent relative dielectric constant. Note that only for
values of $\frac{\epsilon(f)}{\epsilon_{0}}$ as high \ as $10^{8}$, the decay
time$\frac{\epsilon(f)}{\sigma}$ attains values of milliseconds. \ Thus, in
very short periods of time the charges inside the cortex conductor tissues
become screened. Below, we consider that the neural processes (diffusional,
chemical, ....) producing the \ existence of impressed currents, have time
periods very much longer than the screening time constant $\frac{\epsilon
}{\sigma}$.

In particular, in the \ situation where the interior of the region of the
tissue does not contain any impressed currents, the mentioned rapid screening
will enforce that the electric field satisfy the modified Poisson equation
\begin{align}
\overrightarrow{\nabla}\cdot\overrightarrow{J}(\overrightarrow{x}.t)  &
=\overrightarrow{\nabla}\cdot(\sigma(\overrightarrow{x})\overrightarrow{E}%
(\overrightarrow{x},t))\nonumber\\
&  =\overrightarrow{\nabla}\cdot(\sigma(\overrightarrow{x}%
)\overrightarrow{\nabla}\varphi(\overrightarrow{x},t))\nonumber\\
&  =0.
\end{align}

\ On the another hand, considering that the impressed and Ohmic currents
overlap in a given spatial region, and that the impressed current vanishes in
the neighborhood of the boundary; the conservation condition
\begin{align}
\overrightarrow{\nabla}\cdot(\overrightarrow{J}(\overrightarrow{x}%
,t)+J_{imp}(\overrightarrow{x},t)) &  =\overrightarrow{\nabla}\cdot
(\sigma\overrightarrow{\nabla}\varphi(\overrightarrow{x},t)+J_{imp}%
(\overrightarrow{x},t))\nonumber\\
&  =0,
\end{align}
after integration over the interior volume $V_{n}$ of the considered region
leads to
\begin{align}
\int_{S_{n}}d\overrightarrow{s}.\overrightarrow{J}(\overrightarrow{x},t) &
=\frac{d}{dt}\int_{V_{n}}d\overrightarrow{x}(\varrho(\overrightarrow{x}%
,t)+\varrho_{imp}(\overrightarrow{x},t))\nonumber\\
-\frac{\sigma}{\varepsilon}\text{ }Q_{T}(t) &  =\frac{d}{dt}Q_{T}%
(t),\label{Qscreen}%
\end{align}
where, $Q_{T}(t)$ is the total charge contained in the volume $V_{n}$, \ and
the vectorial element of surface $d\overrightarrow{s}$ on the close surface
$S_{n}$ surrounding the neural system, is oriented to the outside of the
region $V_{n}$. The above relation simply expresses that the variation of the
total charge inside any region of the conductor tissue tends to exponentially
vanish with time. \ This vanishing monopole condition for the neural charge
density is a direct consequence of the supposed rapid screening in the
conductor tissue. This situation is usually assumed \ in most of the
discussions regarding measuring of electrical activity in neural systems.
Below we will argue that this conditions could result to be invalid for the
description of the brain current measures being performed today in small
spatial regions \cite{jriera}. \

\section{ \ A model including \ diffusion and active processes}

Let us again underline  that within real neurological systems, impressed currents
created actively by the biological system, can  exist inside the
conductor\ medium  surrounding neuronal membranes and axons, with time
variations in the millisecond range. \ The main observation in support of this
view is that \begin{figure}[h]
\begin{center}
\hspace*{-0.4cm} \includegraphics[width=10.5cm]{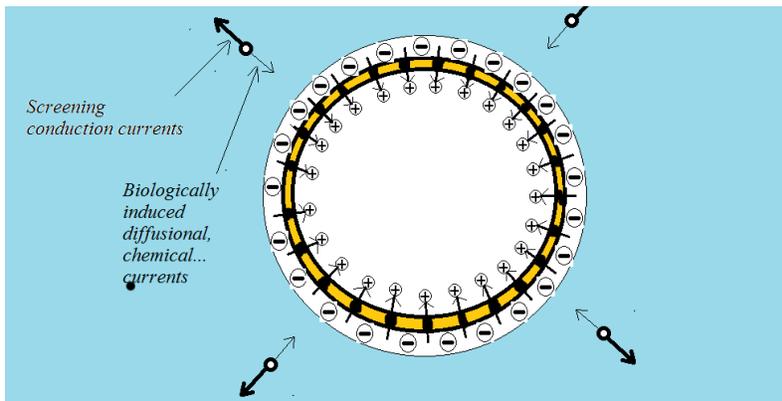}
\end{center}
\caption{ The figure illustrates a neuron in which a flux of positive ions is
passing to the interior of the cell. The existence within the exterior tissue
of a diffusional or chemical processes leading to the flow of ions in the
cell, is indicated by the thin arrows within the exterior conducting tissue.
The picture illustrates the following possibility: Assuming the flow of
positive ions into the cell interior occurs during a time interval, a thin
shell of negative compensating charges will remains close to the membrane
given that the exterior tissue is a conductor sustaining only Ohmic ionic
currents. However, if a diffusive nature is also assumed in the exterior
tissue, such thin shell of negative charges tends to be strongly screened by
external diffusion currents, which are proportional to the gradient of the
density. Thus, positive charges are flowing into the vicinity of the cell up
to some distances which depend on the parameters of the neural tissue. If such
distances are large enough, they can create  measurable signals in the small dimension CSD electrodes,
and then monopolar structures could be detected in experiments. }%
\label{antennae}%
\end{figure}the time intervals for variations of the ionic currents leading to
the neural impulses in axons and dendrites, are in the millisecond range.
Thus, such impressed currents (and probably of diffusive nature) can be
expected to exist  in the close vicinity at the outside of the neurons volume.
A support of this idea comes from  the  results of  Ref. \cite{jriera}, which
are  compatible with a spatial distribution of impressed currents laying
outside the structure of the pyramidal cell to which the reported distribution
is associated. \ The natural blurring of the inverse methods employed to
evaluate these distributions, however, does not allow for a clear definition
of this point.

Figure 1 illustrates a simplified "neuron" defined in the following way. \
Consider a region filled with a homogeneous conductor tissue and within
it, suppose that a biological membrane encloses a region filled by the
referred tissue or by another type of it. Let us suppose that by a given
biological process, an amount of positive ions are assumed to be flowing into
the interior of the membrane, leaving at the exterior a shell in which a
deficit of positive ion exists. \ Note, that if the region is assumed to be
spherical as well as the flux of positive ions, this initial state will show
zero monopole component of the electric field.

\ Let us assume that the conductor tissue is constituted by a density $\rho$
of positive ionic charges which are mobile, and a compensating "jellium" of
negative charges which are assumed to be static in order to simplify the
further discussion by having only one dynamic charge density. \ The dynamics
of the mobile ions are assumed to be given by a generalized diffusion equation
in which the ohmic currents are also included.%

\begin{subequations}
\label{equations}%
\begin{align}
-\nabla^{2}\phi(\overrightarrow{x},t)  &  =\frac{1}{\epsilon}\rho
(\overrightarrow{x},t),\\
0  &  =\frac{\partial\rho(\overrightarrow{x},t)}{\partial t}+\frac{\sigma
}{\epsilon}\rho(\overrightarrow{x},t)-D\text{ }\nabla^{2}\rho
(\overrightarrow{x},t),\\
J_{T}(\overrightarrow{x},t)  &  =-(\sigma\frac{\partial\phi(\overrightarrow{x}%
,t)}{\partial\overrightarrow{x}}+D \frac{\partial\rho(\overrightarrow{x}%
,t)}{\partial\overrightarrow{x}}),\\
\overrightarrow{E}(\overrightarrow{x},t)  &  =-\frac{\partial\phi
(\overrightarrow{x},t)}{\partial\overrightarrow{x}}.
\end{align}

These equations constitute the general definition of the model. In what follows, we will
search for particular explicit solutions of the equations which can be of help
 in discussing the possible relevance of monopolar structures. For the application to theses
  special problems, we will consider boundary conditions in the following simplifying way:

1) \ An explicit and simple form for the initial condition for the ionic
charge density at all the spatial points will adopted for the model.

2) \ After the initial time $t=0,$ the membrane will be assumed to be closed
not allowing the flow of ionic currents through it. \ Mathematically, this
means that the total ionic current $J_{T}$ will be equal to zero at the
membrane. \ This last constraint will avoid the rapid cancelation of the
\ opposite charges existing at both sides of the membrane. \ After the action
potential the neuron will be allowed to return to the standard charge balanced
state.

It should be noted that the real processes in neural systems can be very much
complex, implying, by example, multiple kind of positive and compensating
negative ions. However, in this first discussion we intended to simplify the
analysis by including \ one kind diffusive positive  ionic component with the compensating charge
 assumed to be a "jellium" of fixed negative charges. In addition, a  perfect
 non conductive state of the membranes during periods of time will be assumed, which is clearly
 a reasonable  idealization  for simplifying the finding of explicit solutions.

\ In next sections we will consider the system of equation and its solutions
for the case of a spherical cell. \ However, before considering the spherical
case, let us study the problem for a planar model in which the semi-space
being at the right of a planar membrane is a conductor tissue and the left
half semi-space is assumed to be formed by a medium which does not screen the
entered positive charges and allows them to be mobile within a region near the
membrane. \ This planar system will be qualitatively close to the real
situation in cells, in which there is no possibility of creating charges
within the cell, in order to compensate for the positive ions entering the
cell. Before the initial instant ($t=0$), it will be assumed that the membrane
was \ open for a brief moment in which positive ions entered the left region
forming the assumed initial ionic charge imbalance. After the initial times,
the membrane is assumed to be closed for the flow of ions.

\section{The planar variant of the model}

\ As it was mentioned, in this section we will assume a simplified model of a
cell formed by the membrane, and a right semi-space filled with conducting
tissue. The left semi-space is assumed to be an empty region in which the ions
having entered the intracellular space are mobile but held near the membrane.
\ Therefore, the $x$ coordinate is defined as measuring the distance from a
point within the right semi-space to the membrane, $x>0$. Equations in this
semi-space will have the form\
\end{subequations}
\begin{align}
-\frac{\partial^{2}}{\partial x^{2}}\phi(x,t) &  =\frac{1}{\epsilon}%
\rho(x,t),\\
0 &  =\frac{\partial\rho(x,t)}{\partial t}+\frac{\sigma}{\epsilon}%
\rho(x,t)-D \text{ }\frac{\partial^{2}}{\partial x^{2}}\rho(x,t),\\
\frac{\partial E(x,t)}{\partial x} &  =\frac{1}{\epsilon}\rho(x,t),\\
J_{T}(0^{+},t) &  =-(\sigma\frac{\partial\phi(0^{+},t)}{\partial x}%
+D \frac{\partial\rho(0^{+},t)}{\partial x})=0,\\
J_{T}(x,t) &  =-(\sigma\frac{\partial\phi(x,t)}{\partial x}+D \frac
{\partial\rho(x,t)}{\partial x}),\\
E(x,t) &  =-\frac{\partial\phi(x,t)}{\partial x}.
\end{align}

In the approximation under consideration, the problem is a one dimensional one
in the vicinity of the membrane and the coordinate $x$ has been chosen to be
equal to zero at the membrane position. \ \ The one dimensional problem can be
\ solved by the taking the \ exponential \ Fourier transform with respect to
time. Further, the spatial Fourier Transforms can be employed in order to
expand the \ spatial dependence.

In order to complete the definition of the model, we need only to define a
reasonable functional form for the \ ionic charge densities at the exterior
and interior of the cell. \ \ The densities at the exterior will be specified
as a function of the $x$ \ coordinate at the initial time $\ t=0$ by the
expression
\begin{equation}
\rho(x,t)=\text{ }-\lambda\text{ }x\text{ }\Theta(a-x)\Theta(x)\equiv\rho
_{0}(x),
\end{equation}
where the constant $\lambda$ has dimension of Coulomb over meters to the fourth power, in
order to assure the dimension of \ $\rho$ as Coulombs by cubic  meter,
$\Theta(x)$ is the Heaviside step function and $a$ is the width of the ionic
charge distribution assumed to be created by some active biochemical
processes. \ Note that an arbitrarily chosen strength for the density can be
fixed since  the equations are inhomogeneous, and the solution will be
linearly dependent on the density function at the initial time. We will use
this freedom to choose $\lambda=1$ \ \ C/m$^{4}$ .\ The solution for a specific
charge density can be found by simply multiplying the resulting fields for
$\lambda=1$,  by the correct value of $\ \lambda.$

The solution of the density equations for the chosen boundary condition was
found in the following form. The charge density has the Fourier \ expansion:
\begin{equation}
\rho(x,t)=\int_{0}^{\infty}dq\text{ }c(q,x)\exp(-(D \text{ }q^{2}+\frac{\sigma
}{\epsilon})t)-r_{0}\exp(-\sqrt{\frac{\sigma}{\epsilon D}}x),\label{density}%
\end{equation}
where the function $c(q,t)$ is a linear combination of the sine and cosine
Fourier spatial modes as%
\begin{equation}
c(q,x)=\sqrt{\frac{2}{\pi}}(\alpha\text{ }s(q))\sin(qx)+\beta\text{ }%
(\frac{a^{2}\sigma}{\sqrt{2\pi}(D \,\epsilon \,q^{2}+\sigma)}+c(q))\cos(qx)).
\end{equation}

The $s(q)$ and $c(q)$ \ expansion coefficients have the expressions%
\begin{align}
s(q) &  =\sqrt{\frac{2}{\pi}}\frac{-q\text{ }a\cos(a\text{ }q)+\sin(a\text{
}q)}{(a\text{ }q)^{2}},\\
c(q) &  =\sqrt{\frac{2}{\pi}}\frac{-1+\cos(a\text{ }q)+q\text{ }a\sin(a\text{
}q)}{(a\text{ }q)^{2}}.
\end{align}

In the expression $\alpha+\beta=1$ and $\alpha$ is a constant which was
determined in the process of constructing the solution of the problem. \ The
above formula for the density is a superposition of waves, each one
satisfying the homogeneous electro-diffusion equation for the density. The
coefficients were fixed from the condition of reproducing the initial density
profile at the right of the membrane. However, in this process we had added
and substracted a particular solution of the equation for the density under
the following reasoning. \ As the ionic flow is prohibited by the membrane,
which is assumed to be closed after $t=0$, it is clear that the electric field
near the membrane at its outside will have a constant value defined by Gauss
law. \ It should be noted that in this diffusional case, there can not exist a
surface density of charge, because, in presence of diffusion this will be
imply an infinite diffusional current density which is proportional to the
time gradient of the ion density. \ In addition, the equation for the density
has static modes which vary exponentially \ at infinity. \ This fact suggests
the convenience of adding and subtracting to the found solution a static
solution which directly solves the time independent problem. Since the static
solution exponentially decays at faraway regions, the electric field should
tend to zero at infinity. Therefore, after applying Gauss law, the charge of the system seen at infinite
should be zero. \ This in turns,  allows to determine the amplitude of the static
solution from the condition of having a net charge of opposite sign but equal
in magnitude to the \ positive ionic charge being at the left of the membrane.
This solution corresponds to the last term in (\ref{density}). \ Further, the
subtracted term enforces that the solution minus this substracted term, has a
zero total charge in the whole axis at the beginning of the process. Now,
since the static solution, by construction, has zero flux of particles at all
points, in particular at the membrane, the substracted solution (the first
term in (\ref{density})) should also have zero flux at the membrane. \ \ But,
precisely this condition can be directly satisfied, by choosing the value of
the, up to now indefinite, parameter $\alpha$. \ This point can be understood
after noting that for $\alpha=0$ the expansion for the density is a pure
cosine expansion. Therefore, the total current at the membrane of the solution
vanishes. This property follows because the diffusional current is zero at the
membrane as consequence of fact that the derivative of the density has a sine
Fourier expansion vanishing at $x=0$. Also, since the considered density has
\ zero net charge on the whole axis, it also corresponds to a zero electric
field at the membrane position. \

The expression for the electric field after integrating the density from a
point close to the membrane to an arbitrary one, can be written in the form
\begin{align}
E(x,t) &  =\frac{1}{\epsilon}\sqrt{\frac{2}{\pi}}\int_{0}^{\infty}dq\text{
}{\large (}\alpha\text{ }s(q)\frac{1-\cos(qx)}{q}+\beta\text{ }(\frac
{a^{2}\sigma}{\sqrt{2\pi}(D\,\epsilon \, q^{2}+\sigma)}+c(q))\frac{\sin(qx)}%
{q}{\large )}\exp(-(D \,q^{2}+\frac{\sigma}{\epsilon})t)+\nonumber\\
&  \frac{a^{2}}{2\epsilon}-\frac{r_{0}}{\epsilon}\sqrt{\frac{\epsilon
D}{\sigma}}(1-\exp(-\sqrt{\frac{\sigma}{\epsilon D}}x)).
\end{align}

The value of the electric field at the membrane is given  by Gauss law,
which defines the constant $\ r_{0}$ by
\begin{equation}
E(0^{+},t)=\frac{a^{2}}{2\epsilon}=\frac{r_{0}}{\epsilon}\sqrt{\frac{\epsilon
D}{\sigma}}.
\end{equation}

With the above relations for the density and the electric field the total
ionic current flowing through the membrane at any point $\ x$, can be
evaluated using the formula
\begin{equation}
J_{T}(x,t)=\sigma E(x,t)-D \frac{\partial\rho(x,t)}{\partial x}.
\end{equation}

\subsection{New variables definitions}

Let us discuss the properties of the above solution for the planar case. \ It
can be noted that by a linear transformation of variables, the space
coordinate $x$, the time variable $t$ and the electric field can be redefined
to express the equations in the form%
\begin{align}
0  &  =\frac{\partial\rho^{\ast}(x^{\ast},t^{\ast})}{\partial t^{\ast}}%
+s \,\rho^{\ast}(x^{\ast},t^{\ast})-\text{ }\frac{\partial^{2}}{\partial
x^{\ast2}}\rho^{\ast}(x^{\ast},t^{\ast}),\\
\frac{\partial E^{\ast}(x^{\ast},t^{\ast})}{\partial x^{\ast}}  &  =\rho
^{\ast}(x^{\ast},t^{\ast})\Theta(x^{\ast}),\\
s =\sigma^{\ast}  &  =\frac{\sigma a^{2}}{\epsilon D}.
\end{align}
This form is equivalent to a selection $\ \ \epsilon=1,$ $D=1$ in the original
equations. The starting  values for the variables can be recovered by using the
following relations
\begin{align}
x^{\ast}  &  =\frac{x}{a},\\
t^{\ast}  &  =\frac{D}{a^{2}}t,\\
E^{\ast}  &  =\frac{\epsilon}{a}E,\\
\rho^{\ast}  &  =\rho.
\end{align}

For $N_{a^{+}}$ions in extra cellular medium the diffusion constant
$D_{N_{a^{+}}}=1.33\times10^{-5}$ $cm^{2}/s$ (as reported in references
\cite{QianSejnowski1989,Goodman}) and a charge density size $a=20$ $\ \mu$m,
one unit of the newly defined time means $\frac{a^{2}}{D}=0.3$ \ seconds
\ From now on, the variables to be employed will be those having an asterisk,
but in order to simplify the notation, the asterisk will be omitted beyond
this point.

\subsection{The vanishing conductivity limit}

Let us consider first the case in which the conductivity is assumed to vanish.
This case is interesting, because it allows for the possibility that
the electric field can be nonvanishing at infinity. \ It should be noted that
a finite electric field at infinity and a finite conductivity implies infinite
power dissipation by the system per any finite area of the membrane. This
amount of power can not be delivered by the considered system, which has at
the beginning a finite amount of energy per unit area of the membrane.
\begin{figure}[h]
\begin{center}
Fig.2a)\hspace*{-0.4cm} \includegraphics[width=5.5cm]{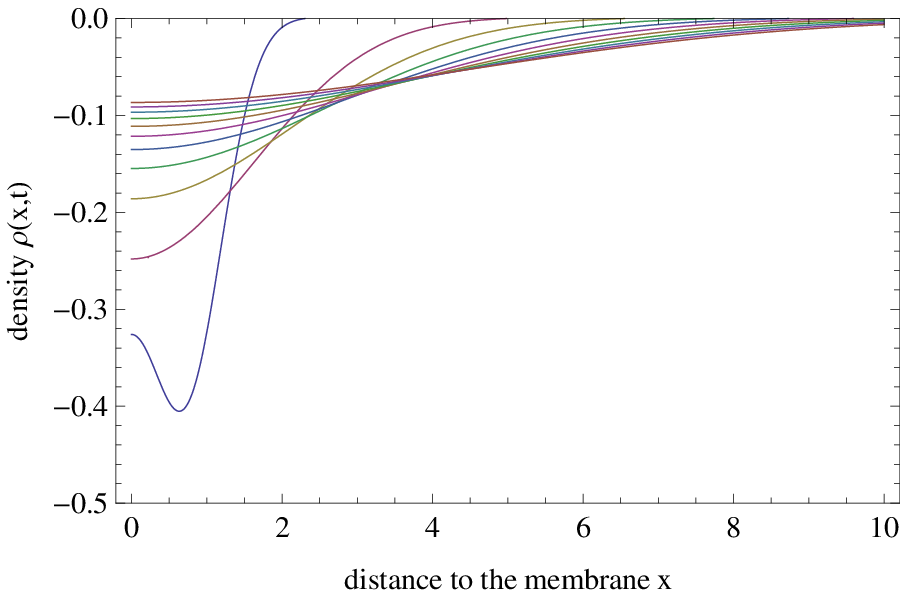}
Fig.2b)\hspace*{-0.4cm} \includegraphics[width=5.5cm]{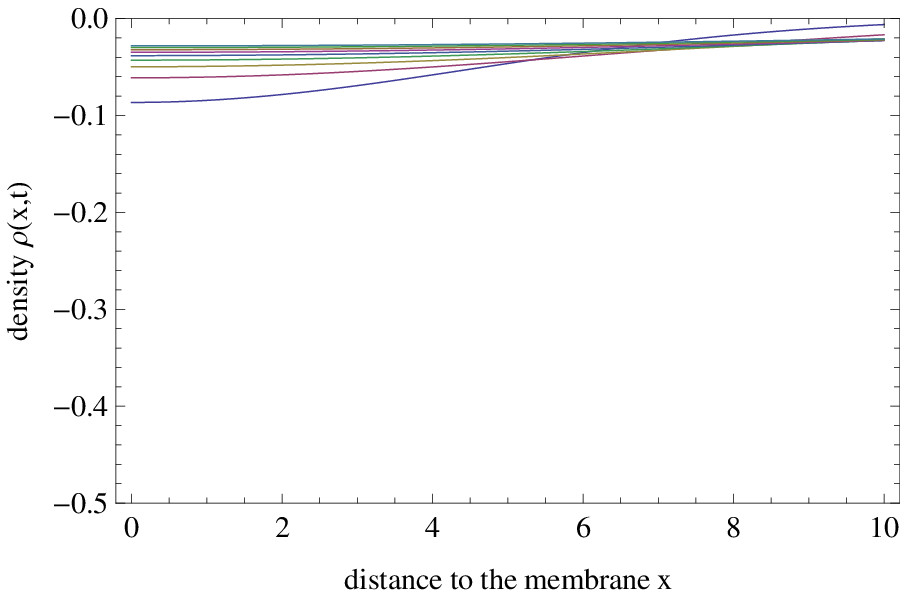}

\end{center}

\begin{center}
\hspace*{-0.4cm} Fig.2c)\includegraphics[width=5.5cm]{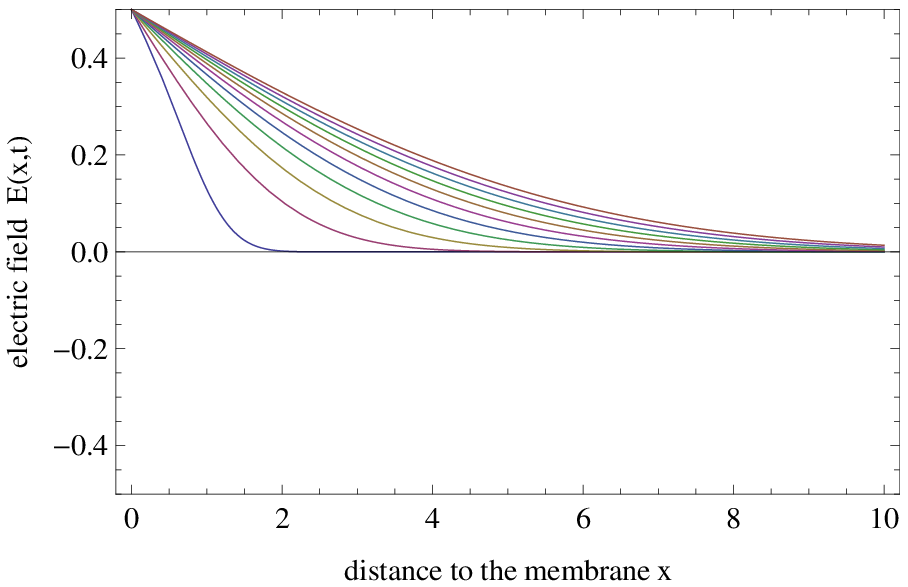}
\hspace*{-0.4cm} Fig.2d)\includegraphics[width=5.5cm]{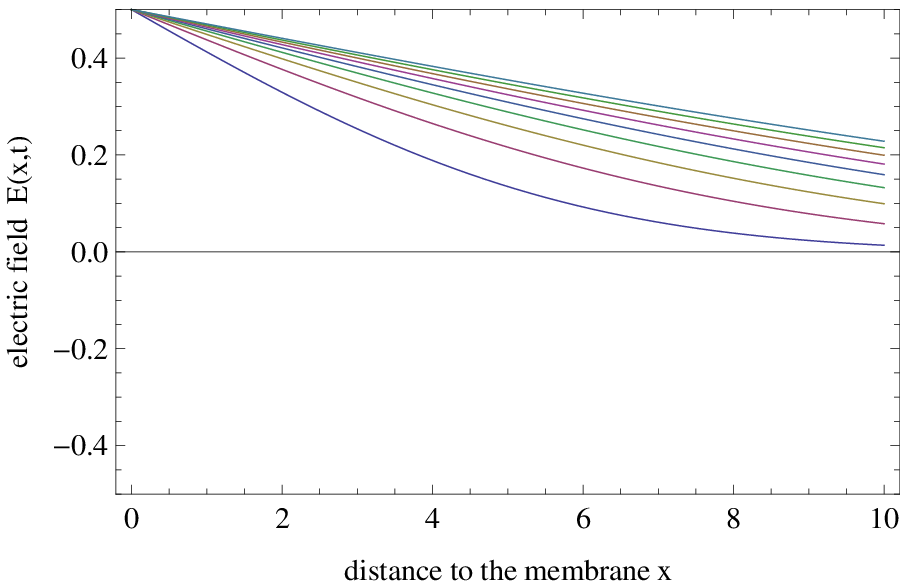}

\end{center}
\caption{ The two top figures shows the plots of the ionic density as functions of the
distance to the membrane for the zero conductivity solution. The top  left
figure contains the plots for the  time values  (0.1,2,3,..,10). In this time
interval the initial peak in the density is damped. The top right  figure
shows the density plots for times in the interval 10 to 100 at larger times
intervals of 10. In this period the negative density at the right of the
membrane propagates to large distances.
The two bottom figures show various plots of the electric field as function of
the distance to the membrane for the zero conductivity case. The bottom  left one
contains the plots for times values (0.1,1,2,3,...,10) ranging. In this time
interval the electric field rapidly decreases with the spatial distances, but
for larger times their values increases. The bottom  right  figure shows the
electric field for times in the interval 10 to 100 at larger times intervals
of 10. In this period the field continues to increase with time. }%
\label{antennae}%
\end{figure}
\ However, if the zero conductivity case shows a monopole in a large finite
volume, being comparable with the volume occupied by a nanometer array of
detectors, then it will directly imply that at least for certain range of
small values of the conductivity, the array could detect electric fields
indicating a monopole component in the measured volume. \ Then, the
explanation of the observed monopoles in Ref. \cite{jriera}, could be
rationalized if the conductivity and diffusion constant of the experimental
system have the appropriate ratio. The question about the experimental values
for this ratio will be discussed in the last sections. \ \

Figure 2a), shows the spatial dependence of the time dependent solution for
ionic charge density as a function of the distance to the membrane for the zero conductivity
case. \ The  curves corresponds to the times values ($0.1,1,2,3,....10)$ in
order to illustrate the starting period of the evolution. Then, figure 2b)
next shows another set of ten curves corresponding to times ranging from $10$
to $100$at intervals of $10$, illustrating the later evolution.   The plots illustrate how the charge
density at the right side of the membrane decays with time. The  curves for
times increasing in figure 2a), show the disappearance of the
initial triangular form of the ionic charge distribution. \ Figure 2b)
illustrates that the diffusion is increasingly smoothing the spatial
dependence of the density at all spatial points; this implies the tendency to
vanish of the current flow at all spatial points.
\ Figures 2c) and 2d), exhibit the values of the electric field as a
function of the distance to the membrane, for the same  corresponding sets of
time values previously defined for figures 2a) and 2b). \ The results
constitute an evidence of the ability of the considered planar model at
vanishing conductivity, of showing a monopole charge densities at faraway
regions from the membrane. \ This is directly implied by the non-vanishing
values of the electric fields even in the large time limit for which the field
tends to be constant in all space as generated by a monopole charge. This net
charge is associated with ions which entered to the cell to the left of the
membrane at the initial instant. This result, in the case of the vanishing
conductivity assumed for the tissue, is allowed by the fact that the system is
able to sustain an electric field in all space, showing a monopole, because
there are not dissipation losses due to a finite electric conductivity. \

\subsection{The finite conductivity limit}

\ Lets us now present results for a non-vanishing value of the conductivity.
We will consider the solution for the parameter value
\[
\sigma^{\ast}=\frac{\sigma a^{2}}{\epsilon D}=0.01,
\]
in order to search for solutions retaining the property that the electric
field shows appreciable values, at distances 10 times the initial size of the
charge distribution $a$. \ The ionic charge density profiles for the same sets
of time values chosen in the past section were evaluated.
\begin{figure}[ptb]
\hspace*{-0.4cm} Fig.3a) \includegraphics[width=5.5cm]{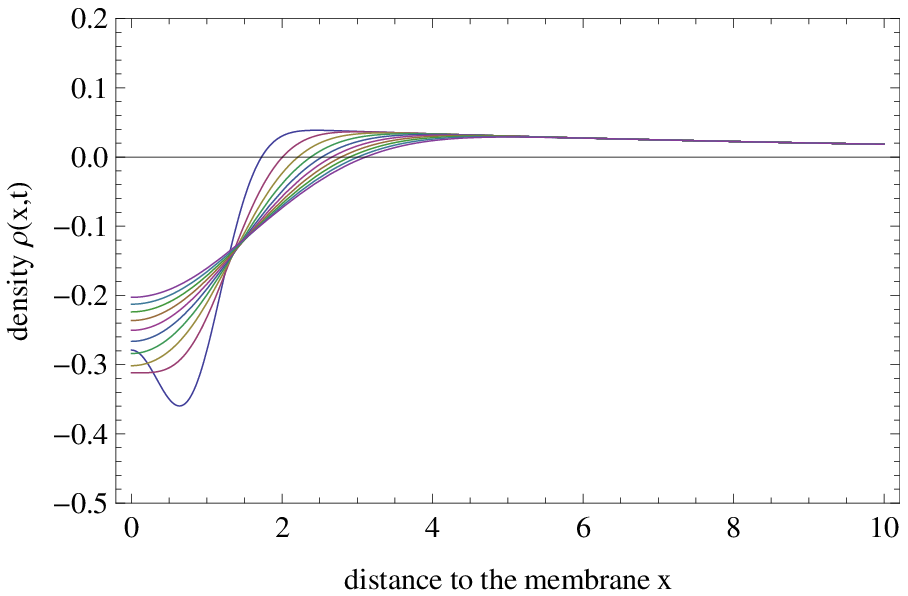}
\hspace*{-0.4cm} Fig.3b)\includegraphics[width=5.5cm]{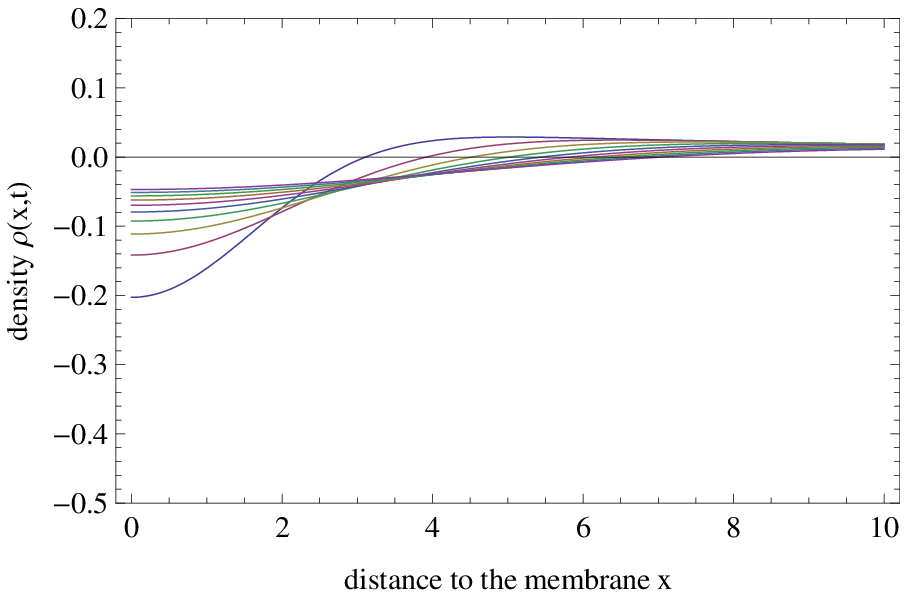}
\hspace*{-0.4cm} Fig.3c)\includegraphics[width=5.5cm]{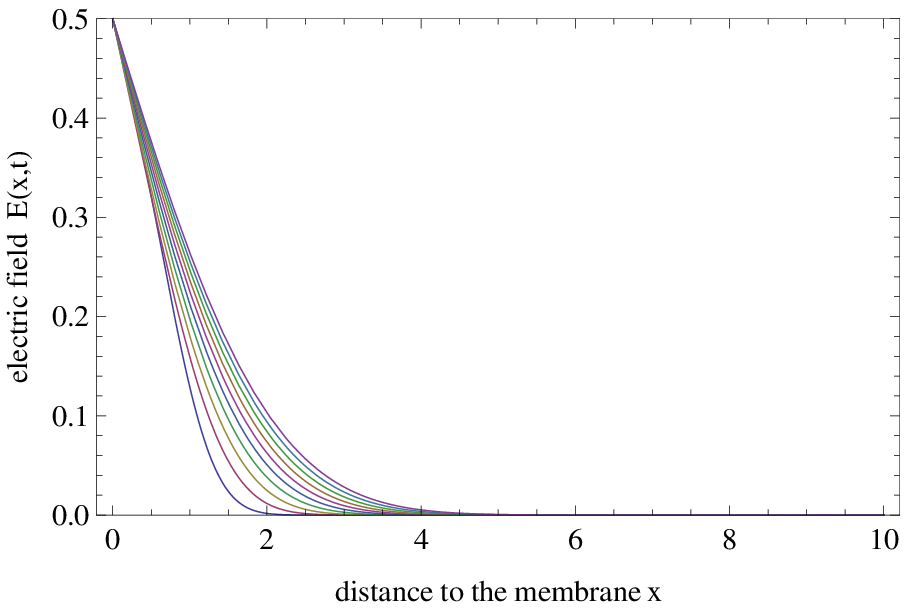}
\hspace*{-0.4cm} Fig.3d)\includegraphics[width=5.5cm]{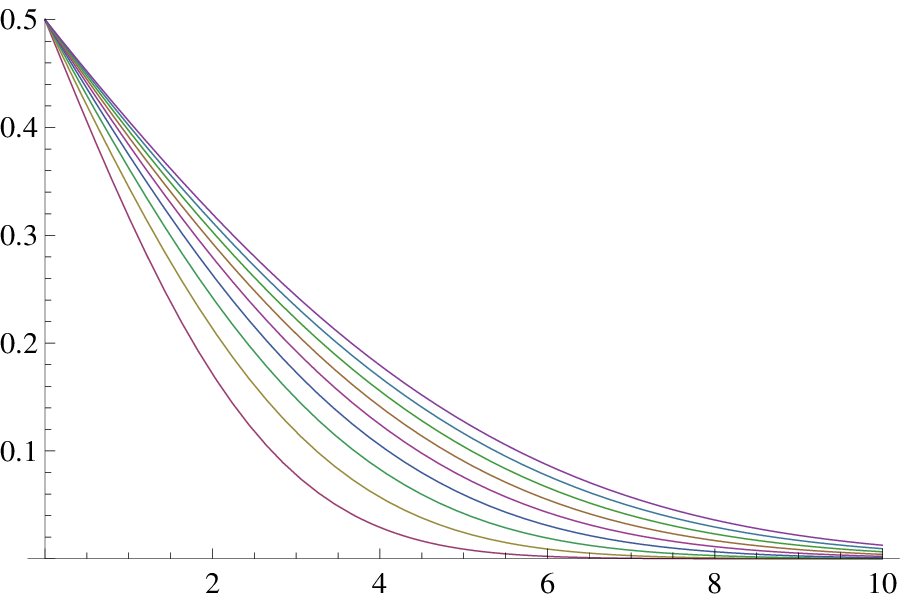}
\caption{ Figures 3a) and 3b) present the results for the time evolution
of the time dependent component of the ionic charge densities at a set of
instants for the finite conductivity solutions. It should be noted that the
time dependent component of the density is the density after subtracting the
previously defined static part. For figure 3a) the times take the values $0.1,1,2,3,...,10$,
for figure 3b) the time varies between $10$ to $100$ at intervals of
$10$.  From
figure 3a) it can be seen the \ charge densities in the beginning, while the
initial triangular distribution is being damped, also tends to rapidly
decrease at large distances.
Figures 3c) and 3d) depicts the electric field spatial
behavior, of the finite conductivity system, for the same three sets of times
instants. The field monotonically decays with the distance to the membrane,
and tends to grow with the time at any fixed point as in the previous zero
conductivity case. However, in the large time limit, the spatial behavior of
the electric fields tends to be exponentially decaying it as should be. For
the parameters selected, it follows that at distances of the order of ten
times the size of the initial charge density, the field (or what is the same,
the monopole that a measuring system of electrodes situated inside the zone
can measure) is an appreciable fraction of the charge trapped with the planar
cell. This behavior leads to the interesting conclusion that at large
distances (nearly ten times the width of the initial ionic density) and large
times, the amount of \ monopole at the membrane measured can be a relatively
large fraction of the charge density of the initial distribution per unit area
of the membrane. }%
\label{antennae}%
\end{figure}

Then, figures 3a) and 3b) present the results for the evolution
of the time dependent component of the ionic charge densities at a set of
instants. It should be recalled that this time dependent component of the
density, is the value after subtracting the previously defined static part.
For figure 3a) the times take the values ($0.1,1,2,3,...,10)$, for figure 3b)
the time varies between $10$ to $100$ at intervals of $10$.  From figure 3a) it can be
seen that the charge densities in the beginning, while the initial triangular
distribution is being damped, also tends to rapidly decrease at large
distances. Figures 3b)  show that at larger times, the density tends
to become homogeneous and the time decay of the field in this phase becomes slower.

The spatial behavior of the electric field is depicted in figures 3c) and
3d). For the same two sets of times instants, the field monotonically decays
with the distance to the membrane and the field tends to grow with time at
fixed spatial points as in the previous zero conductivity case. However, in
the large time limit, the spatial behavior of the electric fields tends to be
exponentially decaying as should be. For the parameters selected, it follows
that at distances of the order of ten times the size of the initial charge
density, the field (or what is the same, the monopole that a measuring system
of electrodes situated inside the zone can measure) is an appreciable fraction
of the charge trapped within the planar cell. This behavior leads to the
interesting conclusion that at large distances (nearly ten times the width a
of the initial ionic density) and longer periods of times, the amount of
\ monopole at the membrane measured can be a relatively large fraction of the
charge density of the initial distribution per unit area of the membrane. As
the conductivity becomes larger this effect diminishes and the behavior at
large times decays exponentially with distances of value $10$ $a.$

The presented results could show relevance for the explanation of the measured
monopoles in reference \cite{jriera}. This is  possible in the case
that the conductivity and diffusion parameters in the tissues containing the
neurons in question show parameter values leading to measurable monopole signals.
It is possible that the 3D situation could enhance the effect, since the diffusional charge
dissipation is enhanced, and then allowing the monopole to be higher in
proportion. \ One important point to note is that, assuming that the mechanism
is valid, the important role taken by diffusion in the problem indicates, that
the non-vanishing monopole measures presented in Ref. \cite{jriera} becomes an
notable effect due to the advanced and reduced size pioneering
experiments done in that work.
Finally one interesting point is that the effective conductivity $\sigma
^{\ast}=\frac{\sigma a^{2}}{\epsilon D}$ includes the dielectric constant in
the denominator, indicating that the high permittivity of the biological
tissues can reduce the values of this quantity.

It can be recalled that the characteristic time $\frac{a^{2}}%
{D},$ when considered for charge distribution dimensions $a=20$ $\mu$m and the
before cited value  extracellular diffusion constant of the ion $N_{a^{+}}$,
takes a value close to $0.3$ s. \ Therefore, after also recalling that for the
cortical tissue the time constant $\frac{\epsilon(f)}{\sigma}$ can show
millisecond values (for high values of the relative dielectric constant
$\frac{\epsilon(f)}{\epsilon_{0}}$ of $10^{8}$), the constant $\sigma^{\ast
}=\frac{\sigma a^{2}}{\epsilon D}$ can  approximately get a value of
$2\times10^{2}.$  Thus, the chosen value in this section of $\sigma^*=0.01$, for illustrative purposes
results to be  much smaller with respect to the values estimated for cortical tissues.

\section{Derivation of the 3D solutions \ from the 1D ones}

In this section we will derive spherically symmetric 3D solutions of the
equations (\ref{equations}) from the 1D solutions obtained in the previous
section. For this purpose, let us rename the one dimensional variable $x$,
giving the distance from the observation point to the planar membrane, in the
way $\ r=x$. \ Next, let us define the new fields $\rho^{d}(r)$ and $\phi
^{d}(r)$ in the forms%
\begin{align}
\rho(x,t)  &  =r\rho^{d}(r,t),\\
\phi(x,t)  &  =r\phi^{d}(r,t).
\end{align}

Then, it follows
\begin{align}
\frac{d}{dx}\rho(x,t)  &  =\rho^{d}(r,t)+r\frac{d}{dr}\rho^{d}(r,t),\\
\frac{d^{2}}{dx^{2}}\rho(x,t)  &  =2\frac{d}{dr}\rho^{d}(r,t)+r\frac{d^{2}%
}{dr^{2}}\rho^{d}(r,t)\nonumber\\
&  =r(\frac{d^{2}}{dr^{2}}+\frac{2}{r}\frac{d}{dr})\rho^{d}(r,t),\nonumber\\
&  =r\text{ }\nabla^{2}\rho^{d}(r,t)
\end{align}
where $\nabla^{2}$ is the 3D Laplacian. In identical form for the electric
potentials follows
\begin{equation}
\frac{d^{2}}{dx^{2}}\phi(x,t)=r\text{ }\nabla^{2}\phi^{d}(r,t).
\end{equation}

The above relations allow us to write the equation of motion of the density
and the potential, as written in the new dimensionless coordinates and times
variables introduced before. The relations become
\begin{align}
\frac{\partial\rho(x,t)}{\partial t}+s\,\rho(x,t)-\frac{\partial^{2}}{\partial
x^{2}}\rho(x,t)  &  =r(\frac{\partial\rho^{d}(r,t)}{\partial t}+s\,\rho
^{d}(r,t)-\nabla^{^{2}}\rho^{d}(r,t))\\
-\frac{\partial^{2}}{\partial x^{2}}\phi(x,t)  &  =r\nabla^{^{2}}\rho
^{d}(r,t).
\end{align}

Therefore, after substituting in the equations, common factors can be
canceled and the equations for the new fields turn to be
\begin{align}
\frac{\partial\rho^{d}(r,t)}{\partial t}+s\,\rho^{d}(r,t)-\nabla^{^{2}}\rho
^{d}(r,t)  &  =0,\\
-\nabla^{^{2}}\phi^{d}(r,t)  &  =\rho^{d}(r,t).
\end{align}

Henceforth, the fields $\rho^{d}$ and $\phi^{d}$ satisfy the same equations
but in the three dimensional space. \ Thus, all the solutions found for the 1D
problem now define corresponding solutions of the 3D problem by performing the
defined changes of variables. It will be helpful to underline some properties
relating the two types of solutions:

1) \ The charge associated to the symmetry point $r=0$ in the 3D solution is
proportional to the linear density of the ionic charge which had passed to the
interior of the cell in the 1 D solution. \ This means that the whole cell in
the 3D solution will be represented by the symmetry point \ $r=0,$ and its
attached net charge.

2) \ A sphere of radial size $a$ defines the region in which initial density
of negative compensating charges were defined. These charges cancels the total
positive charges inside the cell (represented by the physical point $r=0$ in
the 3D solution). \ At the initial time, (t=0), the region defined is
analogous to that defined in the 1D solution where the compensating negative
charge density resides in order to compensate for the positive ions that had
entered the lipid bilayer. This defines the meaning to the parameter $a$.

\section{ Two simple firing models for the Spiny Stellate and Pyramidal cells}

In this section, we will apply the derived 1D and 3D solutions to
qualitatively estimate the ratios between the signals produced outside the
cells by two typical transient processes which we will model to occur in their
vicinity. One of the processes will consider an initial density profile
showing spherical symmetry. Such a configuration can qualitatively resemble
the Spiny Stellate cells action potential showing a weak  monopole moment and zero
multipole moments in the experiments of reference \cite{jriera}. The second
process will be one in which two spherically symmetric configurations, will be
superimposed at two spatial locations separated by a given vector
$\overrightarrow{l}$. These two configurations in this second process will be
assumed to have initial density profiles which are identical but with
oppositely signed charge densities. This model seems to reasonably resemble
asymmetrical cell firing showing a dipole like structure.

\ Its is clear that the detection of a non-vanishing signals in the electric
detector, associated to the spherical spatial symmetry of the field in the
first solution, will be identified as a monopole charge. Thus, in order to
estimate the relative strength of the monopole signal, with respect to second
"dipole" like signal, a tactic can be to compare the relative ratio of the
potentials created at some observation point.

As it was seen from the solution for the one dimensional case, the time
dependent solution of the initial value problems for the ionic density evolves
in a way tending to approach a static exponential solution in which the
internal charge is screened without zero current density. That is, the Ohmic
current density is exactly canceled by the diffusion current, to avoid energy
dissipation. This behavior directly translates to the 3D counterpart
solutions. Therefore, as mentioned before, we will take the values of such
static solutions as a measure of the maximal signal at faraway regions at
which the measuring electrodes will be assumed to be situated. This
corresponds to a situation in which the transient solutions are assumed to
decay in a very short time.

Further, it is assumed that these transient solutions could not be detected by
a small bandwidth electric potential detector, or eventually, that they
constitute the rising transient pontentials leading to the peak action
potential measured. Clearly, the measured decaying time evolution of the
signals in the experiment should be directly attributed to the fact that the
internal charges which passed to the internal region of the cell, will slowly
return to the extra cellular medium, with the help of diffusion processes.
These are recognized to exist, after the membranes becomes closed to the rapid
flux of ions occurring during the neural impulse firings. \ Therefore, the
potential of the spherically symmetric solution at far distances will be
assumed to have the expression
\begin{align}
\phi^{mon}(r) &  =\frac{\phi_{0}}{r}\exp(-\sqrt{\sigma^{\ast}}\,\,r),\\
\sigma^{\ast} &  =\frac{\sigma a^{2}}{\epsilon D}.
\end{align}

At this point it seems convenient to again describe the physical meaning of
this Yukawa field as determined by its derivation within the model. \ Note
that close to the symmetry point $r=0,$ the field is similar to the Coulomb
one for a point charge. This point charge, as interpreted in the 1D variant of
the model, describes the ionic charges that entered the cell through the
membrane, which sits at the $x=0$ coordinate value; which after the change of
variables corresponds to the symmetry point. Thus, the net charge associated
to the Yukawa solution can be naturally interpreted as the charge inside the
cell, which in our model shows vanishing size. The parameter $a$ was defined
as the width of the zone in the 1D model in which the negative charges
compensating the net positive ionic charge was assumed to be sitting at the
initial instant of time. Therefore, after the change of variables defining the
3D solution in terms of the 1D solutions is done, the parameter $a$ also
describes the radius of the sphere at the which the compensating charges are
placed at the beginning of the 3D cell firing being modeled.

The physical process in the above written static Yukawa solution can be
described in the following steps:\newline\noindent1) After the rapid "opening"
of the membrane to the ionic flow, a net amount of positive ionic charges is
assumed to diffuse through it at the x=0 point of the 1 D solution.\newline%
\noindent2) At this initial moment, and after interpreting the $x$ coordinate
of the 1D solution as the radial one for the 3D solution, and scaling the
fields, in the 3 D case, the situation in 3D can be viewed as corresponding to
a net positive point charge concentrated at the symmetry point $r=x=0,$
surrounded by an sphere of radius $a$ filled  of negative charges. These charges
exactly compensate the positive point charge at $r=0.$\newline\noindent3)
\ After that, assuming that the membrane becomes "closed" to the further flow
of ions, the negative charges being within the sphere of radius $a$ can not
follow their natural tendency to enter the cell,  and  tend to diffuse away from
the cell in a process which establishes the Yukawa solutions. The stability of this
solution is assured by the fact that the Ohm currents are exactly canceled by
the diffusion ones, leading to a non dissipative state in which the net
current charge current vanishes.
  \ The tendency to enhance monopole components determined by the inclusion of diffusion processes in the
model for the brain tissue, was seen in the previous sections  in the limit of which the
$\sigma^{\ast}$ parameters becomes small. In this \ case the system tends to
become a dielectric and the Yukawa solutions tends to approach the Coulomb
solution in larger regions. \ Then, if the region is of the order of the
similar size as the one associated to the array of detecting electrodes, the
monopole components can become strong. As noted before, the non-vanishing measured potential determined by this
solution can be interpreted as a monopole due to the spherical symmetry of its
spatial distribution. \

 Let us now consider the second configuration, which
might typically represent less spatially symmetrical processes occurring in
neural systems. The  aim will be to compare its strength with the previous one
showing a monopole. For this purpose, as described before, assume a
superposition of two spherically symmetric solutions, \ associated to two
spherical membranes of  identical sizes, which at the beginning have internal
charges (and associated halos of opposite compensating charges of radius $a$)
of equal magnitude, but different signs. The location of the spherical
symmetry points of the two solutions will be supposed to be displaced in a
vector $\overrightarrow{l},$ whose length is assumed to be of order one. That is,
 corresponding in normal units to the same size of the radius "$a$" . It will be
 also assumed that the two electro-diffusion
processes can be approximately considered as non interacting. The initial
charge density of the system will be less symmetrical than the previously
discussed system. However, due to the transient screening processes the
initial charge densities will be gradually evolving due to the \ diffusion and
ohmic currents. \ The screening of the external charges outside the two points
representing the cells is supposed to occur in short times, the inverse of
these times is much larger than the signal bandwidth of the detector. \ In
this situation, at points $\overrightarrow{r}$ near the detector electrodes,
the potentials created by the two cells systems can be approximately written
by the difference between the two Yukawa potentials created individually by
each cell, in the following form
\begin{align}
\phi^{dip}(r) &  =\frac{\phi_{0}}{|\overrightarrow{r}+\overrightarrow{l}|}%
\exp(-\sqrt{\sigma^{\ast}}\,|\overrightarrow{r}+\overrightarrow{l}%
|)-\frac{\phi_{0}}{r}\exp(-\sqrt{\sigma^{\ast}}r)\nonumber\\
&  \simeq\overrightarrow{l}.\overrightarrow{\nabla}\phi^{m}(r)\nonumber\\
&  =-\overrightarrow{l}.\overrightarrow{n}_{r}(\frac{1}{r}+\sqrt{\sigma^{\ast
}})\phi^{m}(r),
\end{align}

where $\overrightarrow{n}_{r}=\overrightarrow{r}/|\overrightarrow{r}|$, and it
\ has been assumed that at the measuring distances  $r\gg|\overrightarrow{l}%
|.$

Therefore, for this special model system the ratio between the signals at the
detector will have the expression
\begin{equation}
|\frac{\phi^{dip}(r)}{\phi^{m}(r)}|=|\overrightarrow{l}.\overrightarrow{n}%
_{r}(\frac{1}{r}+\sqrt{\sigma^{\ast}})|.
\label{ratio}
\end{equation}
It should be noted that this quantity will be assumed to
correspond to the maximal values of the detected potential at the electrodes,
occurring precisely when the charge density halos of the cells had been
annihilated by the diffusion process outside the cells. This process will
translate the halos's charges to conform the fields around the cells as the
superposition of the Yukawa fields created by the ionic charges retained into
the cells. After this period, the interior ionic charges will be assumed to
slowly escape out from the cell's interior thanks to slow returning diffusion
mechanisms, which are known to allow for this decay in the millisecond region.

Note that in the considered systems of units, one unit of distance is
equivalent to the size parameter $a$ which is of the order of the dimensions
of the regions in which compensating charges appear around the cells. Thus,
since the distance from the two cell system to the detector is at least ten
times  larger than  $a$ (for the measuring experiment in reference
\cite{jriera}) the number $\frac{1}{r}$ is smaller than  the unit. Therefore,
for equal or larger  than the unit  values of $\sigma^{\ast}$ the ratio
between the two signals is determined by the constant $\sqrt{\sigma^{\ast}%
}=\sqrt{\frac{\sigma a^{2}}{\epsilon D}}$.

\ \ The above expression (\ref{ratio}) gives a natural result in the limiting situation when
the conductivity is low and the system is practically an insulator. At this
point the monopole signal is much larger than the dipolar one. This is natural
because this is the limit of good dielectric and poor conductor medium in
which the electric field can exist faraway from the central charges. However,
in the contrary situation in which the conductivity is sufficiently large to
satisfy $\sqrt{\sigma^{\ast}}>>1$, the dipolar contribution becomes larger
than the monopole component. \ In the intermediate case in which $\sigma
^{\ast}$ is of order one, the monopole signal can be of the same magnitude as
the dipolar signal. Therefore, for discussing the relevance of the model being
analyzed it is necessary to study the allowed values of the constant
$\sigma^{\ast}$ for the neural tissues involved in the experiments reported in
reference \cite{jriera}.\ \

\section{The $\phi_{d}$/$\phi_{m}$- experimental ranges}

Let us consider the peak (indicated by bottom arrow in Fig. 2 a) in Ref.
\cite{jriera} for an action potential fired by a pyramidal cell. This figure
illustrates the time dependence of an Action Potential fired by a Pyramidal
Cell (left) and the respective spatial distribution of current sources
(top-right). The absolute values of the monopolar and dipolar electric
currents measured in that work, are $I_{m}=0.015$ $\mu$A and \ $I_{d}=0.070$
$\ \mu$A mm$\,$\ respectively. Thus, these measures indicate a clear deviation
of the system from a quasistatic fully Ohmic conductor behavior of brain
tissues at the micrometer scales. We will assume that the \ instant signaled
with the arrow in the figure at the left, indicates the moments in which
transient diffusion processes lead to potential values at the electrodes,
given by the static Yukawa field generated by the ionic charges concentrated
in the cells. Afterwards, the time dependence of the measures should be
associated with the diminishing of the ionic charges within the cells, due to
slower diffusion mechanisms, which naturally tends to expulse the ionic unbalanced
charges.

Now, it can be noted that the deviation from a purely Ohmic behavior of the
brain tissues implied by these experiences is clearly indicated by equation
(\ref{Qscreen}). It directly determines that the monopole charge in a volume
(whose boundary is fully included in a purely ohmic tissue) shows an exact
exponential time decay of the form $exp(-\frac{\sigma} {\epsilon} \ t)$. This
kind of time evolution is incompatible with the repetitive time dependence
exhibited by the monopole intensities $I_{d}$ measured in reference
\cite{jriera}. Therefore, the experimental results seem to clearly indicate
the limitation of the fully Ohmic nature of the charge transport in brain
tissue. In what follows we will explore the implications of the discussions in
the previous sections in which the role of the diffusion was included in a
simplified model for the neural tissues.

Let us initially examine the prediction of our discussion for the $\frac
{\phi_{d}}{\phi_{m}}$ ratio which can be estimated by the previously derived
equation:%
\begin{equation}
\frac{\phi_{d}(\overrightarrow{r})}{\phi_{m}(\overrightarrow{r})}%
=\frac{\overrightarrow{l}.\overrightarrow{r}}{|\overrightarrow{r}|}(\frac
{1}{|\overrightarrow{r}|}+\sqrt{\sigma^{\ast}}).
\end{equation}

The vector parameter $\overrightarrow{l}$ represents the direction and
distance between the positive and negative charge clouds for the dipolar
current source. Based on the preliminary data for the distribution of the
current densities depicted at the right of figure 2 a) in Ref. \cite{jriera},
we assume this parameter ranges between $0-100\ \mu$m.
Let us assume that $\overrightarrow{l}$ and $\ \overrightarrow{r}$ are
parallel and that the parameter $l=|\overrightarrow{l}|=1$, that is,  chosen to approximately coincide
with $a$ in the original units. This reflects the physical assumption that the separation between
the ionic clouds of different signs should be similar to the size of such
clouds, if they are associated to a common cell. Then, the ratio can be
written in the form%
\begin{equation}
\frac{\phi_{d}(\overrightarrow{r})}{\phi_{m}(\overrightarrow{r})}=(\frac
{1}{|\overrightarrow{r}|}+\sqrt{\sigma^{\ast}}).
\end{equation}

The constant $\sigma^{\ast}=\frac{\sigma a^{2}}{\varepsilon D}$ is defined as
a function of the tissue conductivity $\sigma$, permittivity $\epsilon$, width
of ionic charge distribution $a$ and ionic diffusion coefficient $d$. We used
experimental result for the electric permittivity and conductivity of the
bovine's gray matter reported in Ref. \cite{Gabriel1966}, to find lower and
upper bounds for the conductivity and the permittivity both $0.05\leq
\sigma(\frac{S}{m})\leq0.3$ and $5\times10^{5}\leq\frac{\varepsilon
}{\varepsilon_{0}}\leq8$ $10^{7}$, in the frequency range for the
electrophysiological recordings. \ Note that taking into account these bounds
the time decay constant ($\tau=\frac{\varepsilon}{\sigma}$) for gray matter
could range from to 14.7 $\mu$ s to 14.2 $m$ s. \begin{figure}[h]
\begin{center}
\hspace*{0.4cm} \includegraphics[width=10.5cm]{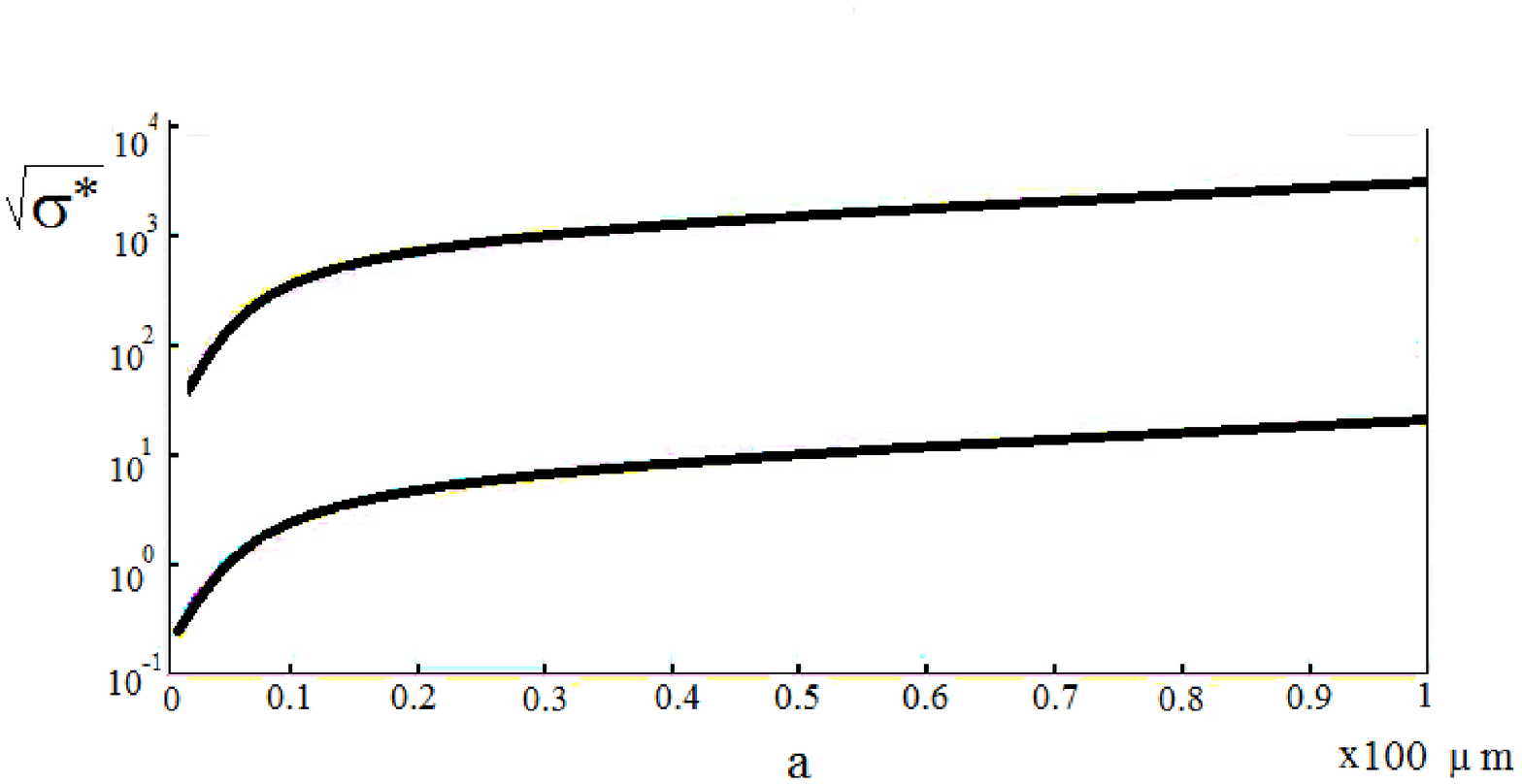}
\end{center}
\caption{ Dependency of $\sqrt{\sigma^{\ast}}$ for the lower (blue) and upper
(red) bounds, as functions of the parameter $a$ measuring the size of the
ionic charge clouds. The upper curve is obtained by using the maximal values
of $\sigma/\epsilon$ estimated for the neural brain tissues in conjunction
with the minimal value of the diffusion constant of the normally involved
ions. On the contrary the minimum value is evaluated by employing the minimal
value of $\sigma/\epsilon$ in common with the maximal value of the ionic
diffusivity. }%
\label{5}%
\end{figure}

The quasi-static approach for the electromagnetic field has been based on the
smallest values of this parameter. However, more complicated phenomena, like
those discussed in this study, could emerge for the largest values of this
decay constant. The lower and upper bounds for the diffusion coefficient were
estimated based on the extracellular ionic profile existing in biological
tissues, with sodium and calcium as the primary ions: $d_{N_{a^{+}}%
}=1.33\times10^{-5}$ $cm^{2}/s$ for the upper bound (See references
\cite{QianSejnowski1989,Goodman}) and $\ d_{C_{a^{2+}}}=0.45\times10^{-5}$
$cm^{2}/s$, for the lower one (See reference \cite{Egelman1999}). The
parameter $a$ on the basis of the experimental data in figure \cite{jriera},
can be in the range of values $1\leq a$ $\leq100$ $(\mu m)$. \begin{figure}[h]
\begin{center}
\hspace*{-0.4cm} \includegraphics[width=10.5cm]{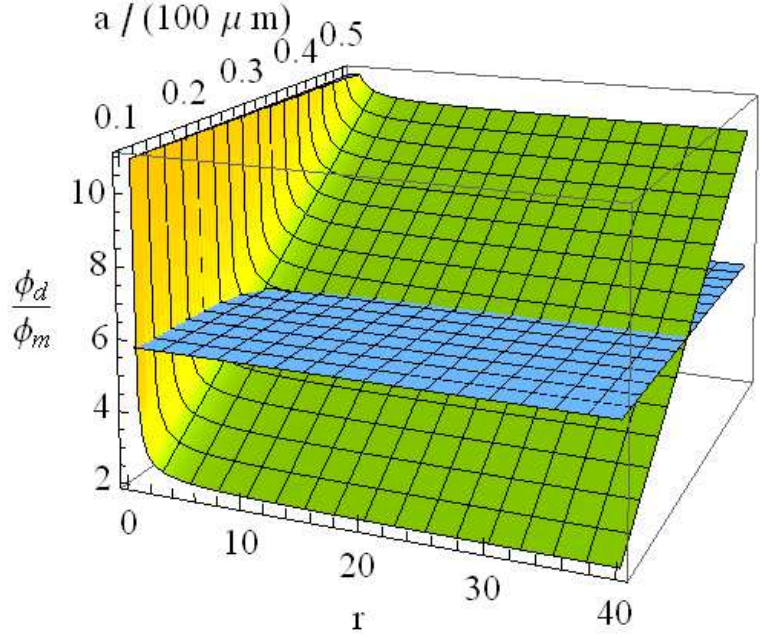}
\end{center}
\caption{ Dependency of $\ \frac{\phi_{d}}{\phi_{m}}$ as a function of the
"radius of the ion clouds" $a$ and the distances $r$ from them to the
measuring electrodes (given in units of $a$). The horizontal plane simply
indicates a typical value close to 6, for the ratio $\ \frac{\phi_{d}}%
{\phi_{m}}$ as estimated from the results for Pyramidal cells of reference
\cite{jriera}. Note that the typical experimental value can be predicted by
the model, for values of $a$ being close to $a=20\ \mu m$. This is valid up to
large distances between the electrodes and the charge clouds of $r=40$, that
is, of $40\ a=800\ \mu m$, after stating that $r$ is measured in units of $a$.
This length is larger than the inter electrode separation of $200\ \mu m$ used
in the experiments reported in reference \cite{jriera}. The value parameter
$l$ was fixed to coincide with $a$, reflecting that the separation between the
ionic clouds of different signs should be similar to the size of these clouds,
if they are associated to a common cell. }%
\label{6}%
\end{figure}Based on these lower/upper bounds, we calculated the dependency of
$\sqrt{\sigma^{\ast}}$ with parameter $a$ (See figure \ref{5}) and of
$\ \frac{\phi_{d}}{\phi_{m}}$ with respect to
$\vert$%
$\overrightarrow{r}|$ \ and $a$ (See figure \ref{6}).

At this point, it should be noted the fact that well defined regions of values
for $\sigma^{\ast}$ and parameter exists which qualitatively match the
experimental results for $\ \frac{\phi_{d}}{\phi_{m}}$ predicted by the
measures reported for Pyramidal cells in reference \cite{jriera}. The plot of
this ratio in figure \ref{6}, clearly provides evidence about the existence of
a wide range of values for $a$ which can produce a qualitative agreement with
the ratios between dipolar and monopolar signals measured in reference
\cite{jriera}, for the Pyramidal cells neuron action potentials. Specifically,
for $\sqrt{\sigma^{\ast}}$ values below the depicted horizontal line traced at
the value 10 in figure \ref{5}, and simultaneously above the lower curve in
this picture, this parameter, which approximately measures the relative
strength between the dipolar and the monopolar signals, shows values between
10 and the unit. These are the magnitudes of the values shown by the
experimental results for the ratios between the dipolar and monopolar signals,
measured in reference \cite{jriera}.

Figure \ref{6} illustrates in more detail
the above comment. It plots the ratio between the dipolar and the monopolar
signal as a function of the parameter $a$ and the variable $r$ (which as
defined before is measured in units of the length $a$). The plot is associated
to the minimal values of $\sigma^{\ast}$ defined by the lower curve in figure
\ref{5}. The value of the parameter $l$ for the plot was fixed to one, which corresponds
with a distance  $a$ in normal units. This assumption is
reflecting that the separation between the ionic clouds of different signs
should be similar to the size of such clouds, if they are associated to a
common cell. The depicted plane indicates a typical value for the ratio of
dipolar to monopolar maximal signals reported in \cite{jriera}, whose value is
approximately 6. Note that for reasonable values of the radius of the initial ionic
clouds of $20\ \mu$ m, the ratio between the dipolar and monopolar signal
becomes close to the experimental values. Interestingly, this occurs at such
large distances between the electrodes and the charge density sources of
$r=40$, which after recalling that $r$ is measured in units of $a$,
corresponds to $800\ \mu$ m. These distances are larger than the values of
$200\ \mu$ m separating the electrodes in the experiments done in reference
\cite{jriera}. Finally, it should also be noted that the experiments for the
Spiny Stellate cells reported in \cite{jriera}, which have an approximate
spherical symmetry, are also in qualitative agreement with the  discussion in
this paper, since for these the measured monopole signal is the only non
vanishing one, and the dipolar and quadrupole components results to be null.

\section*{Summary}

In this work we investigated mechanisms which could generate transient
monopole signals in measuring current source densities. A simple model was
introduced for this purpose. It is concluded that the active and diffusive
natures of the neural biological processes, might determine appreciable
monopole signals in CSD detectors for experiments done at micrometer scales,
due to the ability of the cell membranes and neural tissues to generate ionic
density imbalances in the vicinity of the neurons. Therefore, it is argued
that when both diffusive and Ohmic transport are considered to be present in
neural tissues, electric potential measures in micrometer regions can include
appreciable monopole signals. This might occur for sufficiently small values
of the ratio of the parameter $\frac{\sigma a^{2}}{\epsilon D}$, where
$\sigma$ is the conductivity, $D$ is the diffusion constant and $a$ is the
linear dimension of the ionic charge densities generated by the neural
processes. Possible ranges of magnitudes for these parameters in the
considered experimental studies are estimated. The analysis also predicts
values for the ratio between the dipolar and monopolar signals, that are
similar to those measured in Pyramidal cells in recent experiments reported in
\cite{jriera}. This happens for feasible values of the tissue parameters. As
for the measures in Spiny Stellate cells reported in this same work, the model
also qualitatively reproduces them, by predicting a finite monopolar signal in
combination with vanishing dipolar and quadrupolar signals. The discussion and
results seem to be compatible with the conclusion of a recent work appeared in
Ref. \cite{neuron}. In this paper it is independently stressed the important
role of the active and diffusional nature of the electromagnetic process
occurring in the brain tissues.

The analysis also furnishes a qualitative explanation of the millisecond scales for the measured
processes in the brain tissues. The active  nature of the membrane processes appear in the picture as generating
 maximal signals at the detectors, that afterwards  decay  due to a slower returning of the trapped ions to the extra cellular medium. Both, the neuron membrane firings and the returning ionic currents, then constitute within  the model, the "active' forces  determining the millisecond periods of occurrence of the typical neuron firing  detected in Ref. \cite{jriera}.

Several issues of interest remain to be investigated in connection with the
present analysis. The generalization of the model to more properly include the
presence of various  types of ionic currents is one of the natural
extensions.  Another one is the  study the uniqueness property of the model
equations in connection with the determination of the sources as functions of
the electric potentials. The adoption of the examined equations can  imply
modifications of the algorithms for the evaluation of the sources in terms of
the measured electrode potentials. It is clear that , the  possibility of
"improving" the solution of the inverse problem at such micrometer sized
experiments is suggested by the monopole measurability within the model, this
constitutes a  motivating factor for the further study  of this question.

\section*{Acknowledgments}

J. J. R. would like to express his gratitude to C. B\'edard, A. Destexhe, S. L.
Gratiy, K. H. Pettersen and G. T. Einevoll for helpful conversations and
remarks. A. C. M. wants to acknowledge Jorge Riera, by the kind invitation to
collaborate in the investigation of the subject of this work. The support also
received by A. C. M. from the Caribbean Network on Quantum Mechanics,
Particles and Fields (Net-35) of the ICTP Office of External Activities (OEA)
is also very much acknowledged. Both authors also very much appreciate the
kind revision of the manuscript done by Jos\'{e} Matteo.


\begin{thebibliography}{99}                                                                                               %
\bibitem {jriera}J. J. Riera, T. Ogawa, T. Goto , A. Sumiyoshi, H. Nonaka, A.
Evans, H. Miyakawa and R. Kawashima, J. Neurophysiol. \textbf{108}, 956 (2012).

\bibitem {Bedard&Destexhe}C. B\'edard and A. Destexhe, Phys. Rev. \textbf{E84}, 041909 (2011).

\bibitem {jriera2}J. J. Riera and A. Cabo, J. Neurophysiol. \textbf{109}, 1694 (2013).

\bibitem {Gratiy}S. L. Gratiy, K. H. Pettersen, G. T. Einevoll and A. M. Dale,
J. Neurophysiol. \textbf{109}(6), 1681 (2013).

\bibitem {nunez}P. L. Nunez and R. Srinivasan, \textit{Electric fields of the
brain. The neurophyscis of the EEG,} $2^{nd}$ Edition, p. 158, Oxford
University Press, 2006.

\bibitem {neuron}M. W. Reimann, C. A. Anastassiou, R. Perin, S. L. Hill, H.
Markram and K. Koch, Neuron \textbf{69}, 375 (2013).

\bibitem {diffusion}G. Giebisch, D. C. Tosteson and H. H. Ussing,
\textit{Membrane Transport in Biology. I - Concepts and Model.},
Springer-Verlag Berlin, Heidelberg, 1978.

\bibitem {QianSejnowski1989}N. Qian and T. J. Sejnowski, Biol. Cybern.
\textbf{62}, 1 (1989).

\bibitem {Egelman1999}D. Egelman and P. Pead Montague, Biophys. J.
\textbf{76}, 1856 (1999).

\bibitem {Goodman}J. A. Goodman, C. D. Kroenke and G. L. Bretthorst, Magn.
Reson. in Med. \textbf{53}, 1040 (2005).

\bibitem {Gabriel1966}S. Gabriel, R. W. Lau, C. Gabriel, Phys. Med. Biol.
\textbf{41}, 2251 (1996).
\end{thebibliography}
\end{document}